\documentclass[doublecol,figures]{epl2} 
\usepackage{graphicx}
\usepackage{epsf}
\usepackage{amsfonts}
\usepackage{amsmath}
\usepackage{amssymb}
\usepackage[latin2]{inputenc}
\usepackage{verbatim}

\newcommand{\ci}[1]{c_{\ii #1}}
\newcommand{\cid}[1]{c^\dagger_{\ii{#1}}}
\newcommand{\Bci}[1]{\mathbf{c}_{\ii #1}}
\newcommand{\Bcid}[1]{\mathbf{c}^\dagger_{\ii{#1}}}
\newcommand{\bk}{\mathbf{k}}
\newcommand{\ii}{\mathbf i}

\title{Topological superfluids on a lattice with non-Abelian gauge fields}

\author{A. Kubasiak\inst{1,2} \and P. Massignan\inst{1,3} \and M. Lewenstein\inst{1,4}}
\shortauthor{A. Kubasiak \etal}

\institute{                    
  \inst{1} ICFO-Institut de Ci\`encies Fot\`oniques - Mediterranean Technology Park, 08860 Castelldefels (Barcelona), Spain.\\
  \inst{2} Instytut Fizyki im. M. Smoluchowskiego, Uniwersytet Jagiello\'nski, PL-30059 Krak\'ow, Poland\\
  \inst{3} F\'isica Teorica: Informaci\'o i Processos Qu\`antics, Universitat Aut\`onoma de Barcelona, 08193 Bellaterra, Spain.\\
  \inst{4} ICREA - Instituci\'o Catalana de Recerca i Estudis Avan\c{c}ats, 08010 Barcelona, Spain.
}

\abstract{Two-component fermionic superfluids on a lattice with an external non-Abelian gauge field  give access to a variety of topological phases in presence of a sufficiently large spin imbalance. We address here the important issue of superfluidity breakdown induced by spin imbalance by a self-consistent calculation of the pairing gap, showing which of the predicted phases will be experimentally accessible. We present the full topological phase diagram, and we analyze the connection between Chern numbers and the existence of topologically protected and non-protected edge modes. The Chern numbers are calculated via a very efficient and simple method.}

\begin{document}
\maketitle

\section{Introduction}
The Landau theory of symmetry breaking constitutes the fundamental tool to characterize the different phases of matter.The discovery of the Quantum Hall (QH) and Quantum Spin Hall (QSH) effects showed however, that some physical systems undergo transitions between distinguishable phases without breaking any local symmetry \cite{wen90}. These transitions have a topological character, in the sense that the corresponding "order parameter" is not a local quantity, but rather an integer number describing the system as a whole.

A strong theoretical and experimental effort is now focusing over topological systems, since those are known to contain anyons, quasi-particles which are neither bosons nor fermions. These excitations may have non-Abelian statistics, and are intrinsically robust and protected against decoherence and external perturbations. Non-Abelian anyons have been proposed as the key ingredient for the realization of a topologically-protected quantum computer \cite{kitaev}.
Systems with topological phases include various classes of topological insulators \cite{hasanreview}, the $\nu=5/2$ Fractional Quantum Hall (FQH) state, and two-dimensional (2D) polarized fermionic p-wave superfluids \cite{readgreen,massignan}.

Ultracold gases in optical lattices are at the forefront of current research in the quest for challenging many-body states \cite{BlochRMP}.
 Particularly interesting appear the recent proposals for the realization of artificial gauge fields for neutral atoms \cite{Dalibard10}, since these offer the appealing possibility of realizing \textit{all} possible classes of topological insulators in a single experimental setup  \cite{hasanreview,schnyder}.
 Artificial gauge fields for neutral atoms may be realized via the aid of, as examples, adiabatic Raman passage \cite{dum}, adiabatic control of superpositions of degenerate dark states \cite{ohberg, juzeliunas}, spatially varying Raman coupling \cite{spielmanTheory}, or Raman-induced transitions to auxiliary states in optical lattices \cite{jaksch,osterloh,mueller,demler,dalibard,bermudez}.
Recently Abelian fields have been successfully realized experimentally \cite{spielmanPRL,spielman,gemelke}, and non-Abelian fields may be available soon.

In this paper we study a two-component fermionic gas on a 2D square lattice, with an external non-Abelian gauge field effectively yielding a spin-orbit (SO) coupling in the gas. We take a vector potential of the form $A=(\alpha \sigma_y,\beta \sigma_x)$, where $\sigma_i$ are the Pauli matrices, and $\alpha,\beta$ are independently tunable parameters. This potential can be generated via the techniques proposed in \cite{bermudez,rashba,Dalibard10}. In the absence of interactions, the excitations of this system include massless Dirac fermions, and the  associated Dirac cones may be tuned to be asymmetric in the $x$ and $y$ directions \cite{goldman}. In presence of an additional Abelian field, the ground state is the so called squeezed Laughlin state, and the system exhibits anomalous Integer QH physics \cite{goldman}.
 
Interactions are fundamental in order to access both the superfluid regime and the FQH physics. We study here the case of attractive interactions, which at low temperatures naturally yield BCS superfluid pairing.
In presence of a strong external magnetic field the superfluid enters a topological phase characterized by anyonic excitations with non-Abelian statistics \cite{sato, saujayshort}. The corresponding Hamiltonian is unitary equivalent to the one of a p-wave superfluid \cite{sato}.
Nonetheless, the imbalance needed to observe a phase transition towards a topological state is larger than the one predicted to yield pair breaking, the so called Chandrasekar-Clogston (CC) limit \cite{cc}.

We investigate here this important and still unaddressed question: is a topological phase accessible at all in ultracold s-wave superfluids?  As it is shown in the following, we find a positive answer to this question: the spin-orbit coupling makes a superfluid sufficiently stable to enter various topological phases.
 We continue by  presenting the full topological phase diagram, calculated as a function of the lattice filling and the strength of the spin-flipping couplings along the $x$ and $y$ directions.
  The phase diagram is characterized in terms of the corresponding Chern numbers (CN), which we obtain via a peculiarly simple and efficient method.
Topological phases may be efficiently discussed in terms of the symmetry classes introduced by Altland and Zirnabuer \cite{altland97}. In our system Time-Reversal and Spin-Rotation invariances are destroyed by the Zeeman and SO terms, and as a consequence our Hamiltonian belongs to most general symmetry class D.
The periodic classification of topological insulators \cite{schnyder,hasanreview} predicts that the topological phases of this Hamiltonian in 2D are indexed in terms of a $\mathbb{Z}$ number. 
As we will see, phases with CN=$0,\pm1,\pm2$ are accessible with ultracold fermionic gases in presence of the strong SO coupling provided by a synthetic gauge field.
We study different classes of edge modes in this system, evidencing their strict relation with the Chern numbers.
In analogy with the QSH effect, we also find that the topological protection of the edge modes is destroyed when pairs of states with same spin appear on the same edge. Topological protection of these modes is therefore characterized in terms of the $\mathbb{Z}_2$ index given by the parity of CN. 
\section{Description of the system}

We consider a spin-imbalanced mixture of two fermionic species with interspecies attractive interaction. Particles are free to move in a 2D square lattice in presence of an external non-Abelian field.
Introducing $\Bcid{}=(\cid{\uparrow},\cid{\downarrow})$, the hopping contribution may be written as
\begin{equation}
\mathcal{H}_{\text{KIN+SO}}=-\text{t}\sum_{\ii}\sum_{\hat{e}=\hat{x},\hat{y}}
\left(\Bcid{+\hat{e}}\gamma_{\ii\rightarrow \ii+\hat{e}}\Bci{}+\mathrm{h.c.}\right)
\end{equation}
The hopping terms are defined as:
\begin{equation}
\gamma_{\ii\rightarrow \ii+\hat x}=e^{- i\sigma_y \alpha}\qquad \qquad\gamma_{\ii\rightarrow \ii+\hat y}=e^{ i\sigma_x \beta}
\end{equation}
Using the identity $e^{i\mathcal{A}\sigma_j}={\bf 1}_{2x2}\cos{\mathcal A}+i\sigma_j \sin{\mathcal A}$, it is easy to see that the spin-conserving [spin-flipping] hopping terms in the $x$ and $y$ directions are proportional respectively to $\cos(\alpha)$ and $\cos(\beta)$ [$\sin(\alpha)$ and $\sin(\beta)$]. Adding the chemical potential, Zeeman and interaction terms, we obtain the complete Hamiltonian:
\begin{multline}
\label{completeHamiltonian}
\mathcal{H}=\mathcal{H}_{\text{KIN+SO}}
+\sum_\ii\Delta_\ii\left[\cid{\uparrow}\cid{\downarrow}+\ci{\downarrow}\ci{\uparrow}\right]\\
-\sum_{\ii}\left[(\mu+h)\cid{\uparrow}\ci{\uparrow}+(\mu-h)\cid{\downarrow}\ci{\downarrow}\right].
\end{multline}
Here $V>0$ is the attraction strength, the BCS s-wave pairing $\Delta_\ii=-V \langle\ci{\downarrow}\ci{\uparrow}\rangle$ has been taken real, and a constant term has been neglected.
Due to the mean-field character of the BCS approximation, it is important to realize that the pairing strength $\Delta_\ii$ is not an external parameter which is tunable at will, but it is a quantity that has to be calculated self-consistently. This point will be analyzed in detail in the next Section. 

In presence of periodic boundary conditions, the pairing gap may be taken as constant throughout the system ($\Delta_\ii=\Delta$) and the Hamiltonian can be easily diagonalized \cite{sato}. We introduce $\Psi_\bk^\dagger=(c^\dagger_{\bk,\uparrow},c^\dagger_{\bk,\downarrow},c_{-\bk,\uparrow},c_{-\bk,\downarrow})$ with 
$\mathbf{c^\dagger_k}=V^{-1/2}\sum_\ii e^{i\bk\cdot \ii}\mathbf{c^\dagger_\ii}$, and obtain
\begin{equation}
\mathcal{H}=\frac{1}{2}\sum_\bk \Psi^\dagger_\bk \mathcal{H}_\bk \Psi_\bk,
\end{equation}
\begin{equation}
\mathcal{H}_\bk=\left(\!\!\!\!
					\begin{tabular}{cc}
					$\epsilon_\bk-h\sigma_z+{\bf g_k}\boldsymbol\cdot\boldsymbol\sigma$ & $i\Delta\sigma_y$\\
					$-i\Delta\sigma_y$ & $-\epsilon_\bk+h\sigma_z+{\bf g_k}\boldsymbol\cdot\boldsymbol\sigma^*$
						\end{tabular}
					\!\!\!\!\right)
\end{equation}
where $\boldsymbol\sigma = (\sigma_x,\sigma_y)$. The spin-conserving dispersion is given by
\newline
\begin{equation}
\epsilon_\bk = -2\text{t}(\cos{\alpha}\cos{k_x}+\cos{\beta}\cos{k_y})-\mu,
\end{equation}
while the spin-flipping term arising from SO coupling reads
\begin{equation}
{\bf g^\dagger_k}=2t\left(\sin{\beta}\sin{k_y},-\sin{\alpha}\sin{k_x}\right).
\end{equation}

The eigenvalues of this Hamiltonian are
\begin{equation}
\lambda^2_\bk=\epsilon^2_\bk+h^2+|{\bf g_k}|^2+\Delta^2 \pm 2\sqrt{h^2(\epsilon^2_\bk+\Delta^2)+\epsilon^2_\bk |{\bf g_k}|^2}.
\label{eigenV}
\end{equation}
In presence of pairing ($\Delta\neq 0$), the spectrum is generically gapped, but the gap may close at the four distinguishable momenta which yield ${\bf g_k}_0=0$, i.e.,  \mbox{$\bk_0\in\{(0,0),(0,\pi),(\pi,0),(\pi,\pi)\}$}. The four values of the spin-imbalance at which the gap closes are given by
\begin{equation}
h_{\bk_0}=\sqrt{\epsilon_{\bk_0}^2+\Delta^2}.
\label{gapClosing}
\end{equation}
Transitions between phases with different topological properties may only happen at gap-closing points. Since the balanced superfluid ($h=0$) is known to be topologically trivial, the latter condition states that to access a topological phase one has to consider imbalances satisfying the condition $h>\mathrm{min}_{\bk_0} (h_{\bk_0})$.

\section{Self-consistent calculation of the pairing gap $\Delta$}
\label{selfConsistent}
It has been recently pointed out \cite{sato,saujayshort,longPapers} that topologically non-trivial phases should appear in fermionic s-wave superfluids in presence of both SO coupling and spin-imbalance.
In these works, the pairing gap was introduced as an external parameter tunable at will.
This assumption should be taken with care at large spin-imbalance, since it is well known that superfluidity breaks down when the Zeeman energy (proportional to $h$) becomes large as compared to $\Delta_0$, the pairing gap  at $h=0$ and $T=0$. There, the free energy of the paired state becomes larger than the one of the normal $(\Delta=0)$ state, and the system presents a first order phase transition from superfluid to normal. In absence of SO coupling ($\alpha=\beta=0$) and in the continuum, the critical value is analytic and given by the so called Chandrasekar-Clogston (CC) limit, $h_\mathrm{CC}=\Delta_0/\sqrt{2}$ \cite{cc}. It may be seen from eq.~(\ref{gapClosing}) that $h_{\bk_0}>h_{CC}$ for every $\bk_0$, a condition that excludes the possibility of having topological phases for fermionic s-wave superfluids in the absence of SO coupling. 

\begin{figure}
\includegraphics[width=0.8\columnwidth]{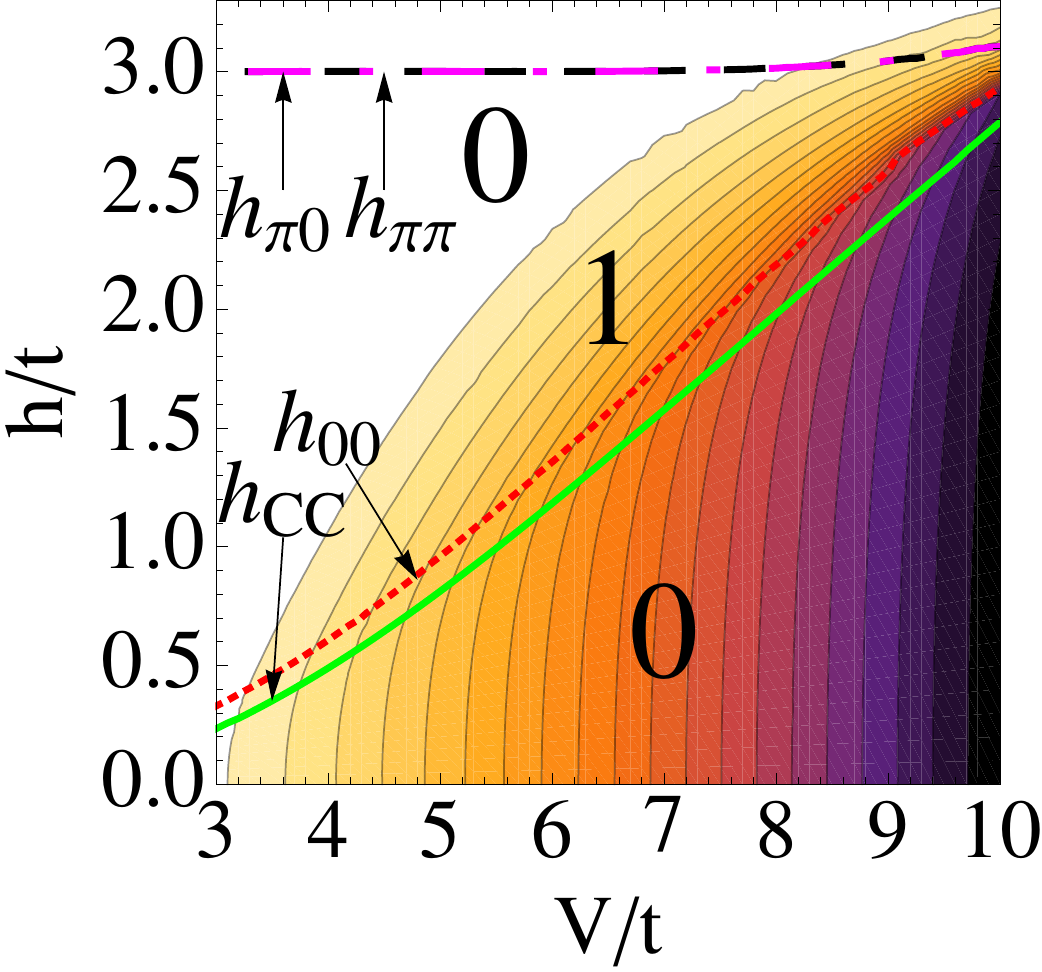}
\includegraphics[width=0.05\columnwidth]{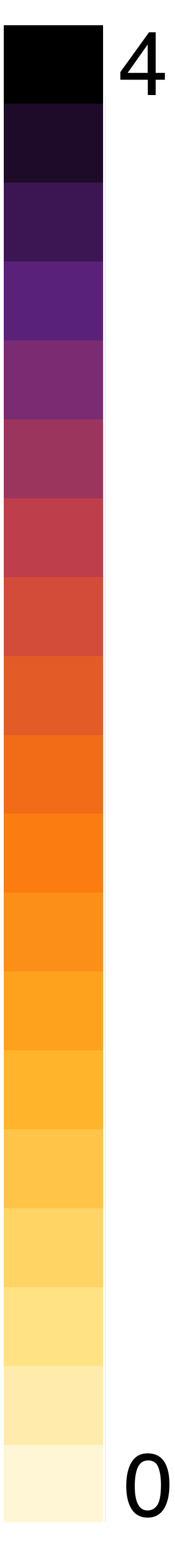}\\
\includegraphics[width=0.8\columnwidth]{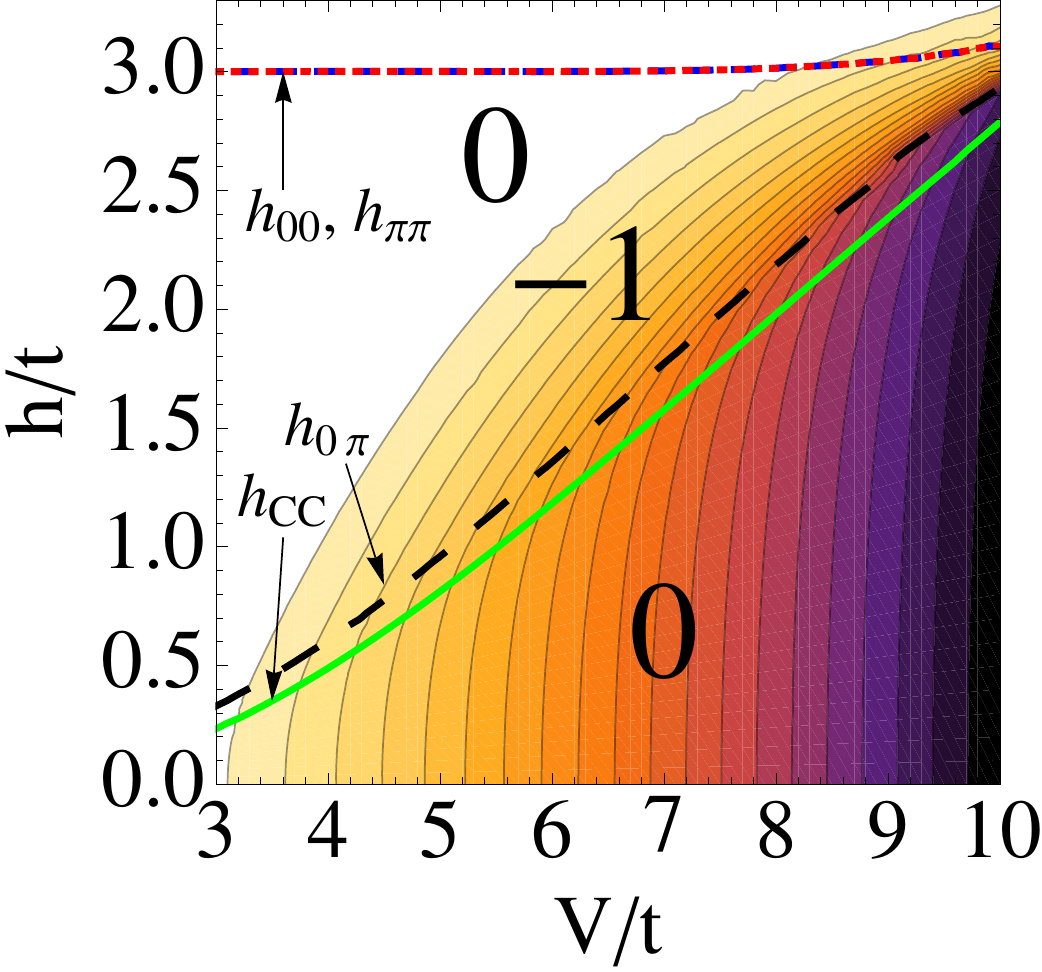}
\includegraphics[width=0.05\columnwidth]{fig1legend.pdf}\\
\includegraphics[width=0.8\columnwidth]{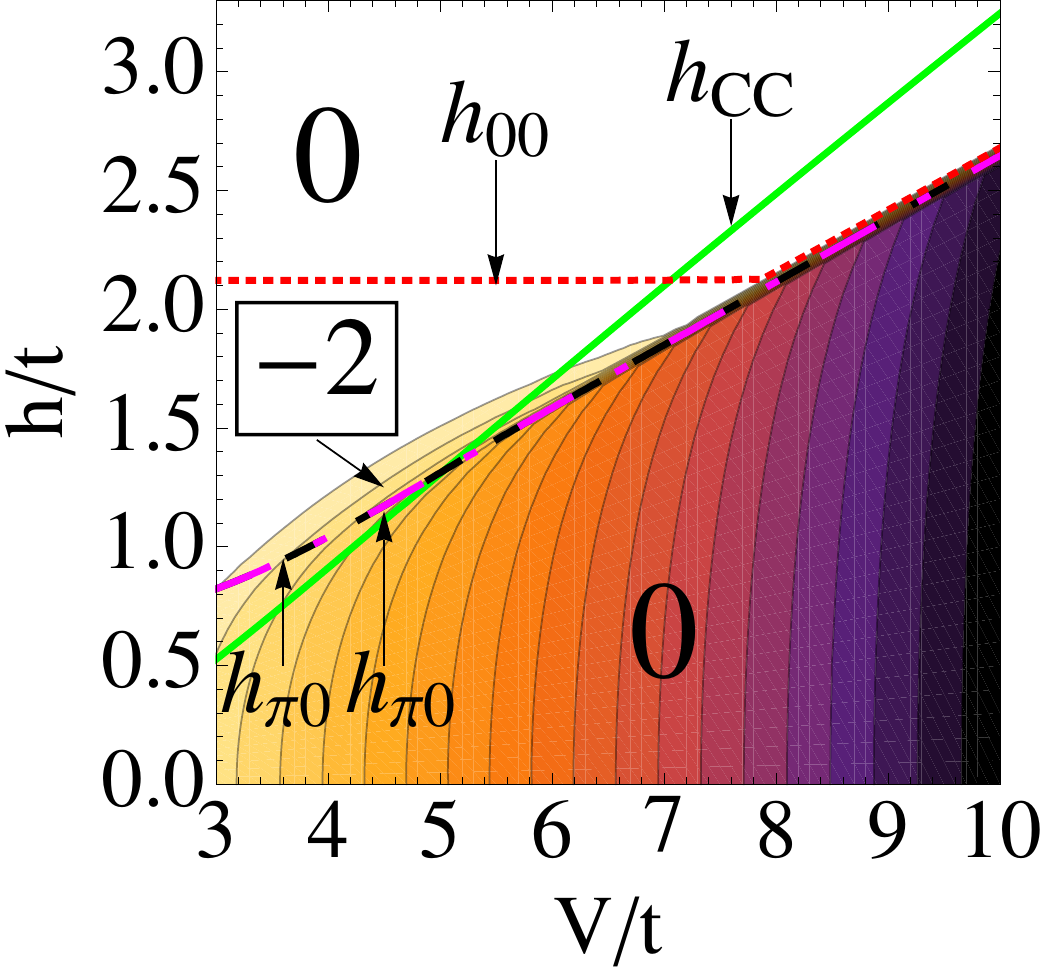}
\includegraphics[width=0.05\columnwidth]{fig1legend.pdf}\\
\caption{
Pairing gap $|\Delta|/t$ as a function of imbalance $h$ and attraction strength $V$.
The lines represent respectively the CC limit $h_\mathrm{CC}=\Delta_0/\sqrt{2}$ (continuous), and the topological boundaries $h_{00}$ (dotted), $h_{0\pi}$ (wide dashed), $h_{\pi 0}$ (dot-dashed) and $h_{\pi\pi}$ (fine dashed), as defined in eq.~(\ref{gapClosing}).
The gauge field strengths and imbalance here are $\alpha=\pi/4$, and $(\beta,\mu)=(\pi/4,-3t)$ (top), $(3\pi/4,-3t)$ (center), and $(\pi/4,-\sqrt{2}t/2)$ (bottom). The numbers inside the plots give the CNs of the most accessible phases. The Chern number vanishes when $\Delta=0$, since a gas in the normal phase is always in a trivial topological state.
}
\label{selfConsistentDelta}
\end{figure}

The self-consistent calculation of the pairing gap $\Delta$ proceeds via the minimization of the free energy
\begin{equation}
F=\frac{N\Delta^2}{V}+\sum_\bk\left[\epsilon_\bk-\frac{k_{\mathrm B}T}{2}\sum_{i=1}^4 \ln\left(1+\mathrm{e}^{-\lambda_{\bk,i}/k_{\mathrm B}T}\right)\right].
\label{freeEnergy}
\end{equation}
Here $\lambda_{\bk,i}$ $(i=1,\ldots,4)$ are the four solutions of eq.~(\ref{eigenV}), and $N$ is the number of lattice sites.
 In the absence of SO coupling $(\alpha=\beta=0$) and in the continuum limit, by solving eq.~(\ref{freeEnergy}) we find that $\Delta$ drops abruptly to zero at $h=h_\mathrm{CC}$ in agreement with the CC limit.
The results of this calculation in presence of SO on a lattice are shown in fig.~\ref{selfConsistentDelta}, where we plot the magnitude of $\Delta$ as a function of the spin imbalance $h$ and the attraction strength $V$.
 As one cranks up the SO coupling, the sharp phase transition gets smoothened \cite{frigeri}, and we find that values of $\Delta \gtrsim t$ are easily reachable in the regions where $h>\mathrm{min}_{\bk_0} (h_{\bk_0})$. Largest gaps are obtained when the spin-conserving and spin-flipping tunneling energies have comparable magnitudes, i.e., for $|\tan(\alpha)|\sim|\tan(\beta)|\sim 1$. As we will see in the following section, all phases with $|$CN$|\leq 2$ may be reached  at $h\gtrsim \mathrm{min}_{\bk_0} (h_{\bk_0})$ for particular values of $\alpha,\beta,$ and $\mu$. In fig.~\ref{selfConsistentDelta}, this imbalance yields respectively (from top to bottom) CN=1,-1,-2.
The superfluid is instead always unstable due to pairing breakdown at imbalances  much larger than $\mathrm{min}_{\bk_0} (h_{\bk_0})$.

\section{Topological phase diagram}

Our Hamiltonian has built-in particle-hole symmetry, i.e. its eigenvalues come in $\pm$ pairs. The topological state of this system may then be characterized  in terms of the bulk Chern number (CN) of the upper band of the spectrum.
The latter is a $\mathbb{Z}$ topological invariant which may change only when the spectral gap between two bands closes. In addition, the sum of CNs associated to an isolate set of bands is conserved before and after the band touching.
 We calculate the upper band's CN following the elegant method proposed by Bellissard in ref.~\cite{bellissard}. Let's assume that the gap closes at point $\bk_0$ as $h$ crosses the critical value $h_{\bk_0}$. The 
Hamiltonian $\mathcal{H}$ has then two eigenvectors $|\psi_0^{(1)}\rangle,|\psi_0^{(2)}\rangle$ with vanishing eigenvalue. In the neighborhood of $(\bk_0,h_{\bk_0})$, we may approximate the Hamiltonian with an effective $2\times 2$ matrix obtained by projecting $\mathcal{H}$ on  $|\psi_0^{(1)}\rangle,|\psi_0^{(2)}\rangle$. Using the Pauli matrices $\vec\sigma = (\sigma_x,\sigma_y,\sigma_z)$, the effective Hamiltonian may be written as
\begin{equation}
\mathcal{H}_\mathrm{eff}(\bk,h)=E(\bk,h)+\vec{\sigma}\cdot\vec{f}(\bk,h).
\label{Heff}
\end{equation}
If the spectrum features a linear Dirac cone at $(\bk_0,h_{\bk_0})$, the Jacobian matrix $J_{\vec{f}}$ has a non-zero determinant at this point.
In this case, as $h$ increases above $h_{\bk_0}$, the change in Chern number of the upper band is given by the sign of the determinant itself. This quantity is called the Berry index of the gap-closing point. More generally, if at a given value $\tilde{h}$ the spectrum features multiple Dirac cones centered in $\bk_0^{(1)},\bk_0^{(2)},\ldots$, the change in Chern number at $h=\tilde{h}$ is given by the sum of the Berry indices at all the touching points,
\begin{equation}
\Delta \mathrm{CN}(\tilde{h})=\sum_i \mathrm{sign}\{\mathrm{det}[J_{\vec{f}}(\tilde{h},\bk_0^{(i)})]\}.
\label{chernFormula}
\end{equation}
From the latter formula, it is clear that each Dirac cone changes the Chern number of a band by $\pm 1$.
If the determinant of the Jacobian at a given gap-closing point vanishes, the Chern number of a band changes there instead by 0, or $\pm 2$ \cite{bellissard}. The latter case is however rather pathological. In the present context, it is verified only when either $\alpha$ or $\beta$ equal integer multiples of $\pi/2$.

\begin{figure*}
\includegraphics[width=2\columnwidth]{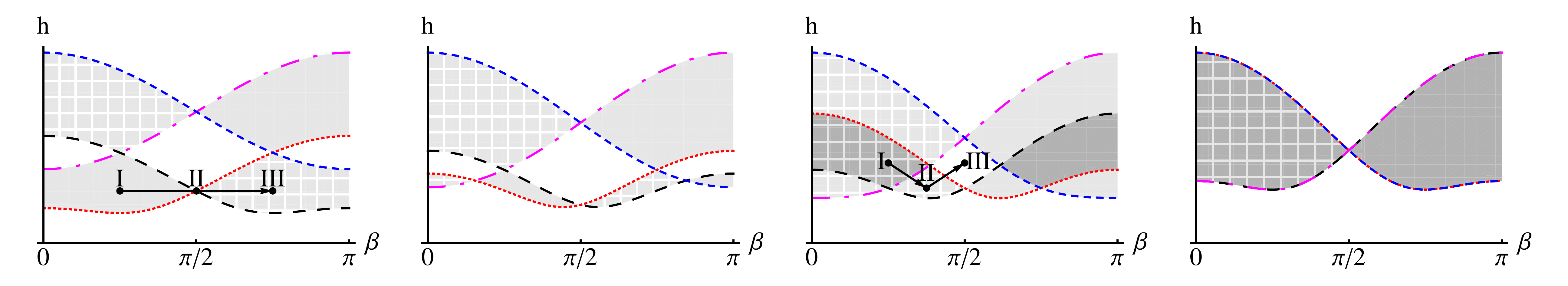}
\caption{
Topological phase diagram of the system as a function of $\beta$, the gauge field strength in the $y$-direction. Here we have set $\alpha=\pi/4$ and $\Delta=t$.
From left to right, the chemical potential is $\mu/(\sqrt{2}t)=-2$ (a),$-1.25$ (b),$-0.5$ (c), 0 (d).
The lines depict the critical imbalances $h_{00}$ (dotted), $h_{0\pi}$ (wide dashed), $h_{\pi 0}$ (dot-dashed), $h_{\pi\pi}$ (fine dashed). The phases are indexed by the Chern numbers calculated from eq.~(\ref{chernFormula}): CN=0 (white), $|$CN$|$=1 (light gray), $|$CN$|$=2 (dark gray), and the meshed phases are the ones with negative CN. The labeled dots in the first and third plots mark the imbalances plotted in figs.~\ref{standardSpectrum} and \ref{invertedSpectrum}.
}
\label{topPhaseDiag}
\end{figure*}

In order to investigate the topological character of various phases, we show in fig.~\ref{topPhaseDiag} the curves corresponding to the four critical imbalances $h_{\bk_0}$  defined in eq.~(\ref{gapClosing}). The curves are plotted as a function of $\beta$ for fixed $\alpha=\pi/4$, and for four increasing values of the chemical potential $\mu$ (from left to right). We consider only negative chemical potentials, since the Hamiltonian has built-in particle-hole symmetry, and reversing the sign of $\mu$ produces symmetric phases to the ones found here.

At small imbalances an s-wave superfluid is topologically trivial for any $\mu$, and CN=0.
 Each time the imbalance $h$ crosses one of the critical curves $h_{\bk_0}$, the gap between the bands closes at the corresponding momentum $\bk_0$, and a topological phase transition occurs.
At low lattice fillings, we find phases characterized by Chern numbers CN=-1,0,1.

For higher lattice filling there appear phases with Chern numbers as large as $\pm 2$.
As we will see in the following section, where we will discuss edge states, the phases with CN=-1,1 are topologically protected, while the phases with CN=-2,0,2 are not protected.
The topological protection of the system is therefore characterized by the $\mathbb{Z}_2$ number given by mod$_2$(CN), in analogy with the theory of the Quantum Spin Hall effect \cite{spinhall}.
At half-filling, $\mu=0$, the phase diagram is composed of regions with CN=-2,0,2, and is therefore not topological for any $\alpha,\beta$, and $h$.

As we have seen in the previous Section, not all regions of the phase diagram plotted in fig.~\ref{topPhaseDiag} are accessible due to pair-breaking. Nonetheless, phases with equal CN are topologically equivalent, and we have shown that phases with all CNs (-2,-1,0,1,2) may be realized with sufficiently large $\Delta(\sim t)$ when $h \gtrsim \mathrm{min}_{\bk_0} (|h_{\bk_0}|)$. 

\section{Spectrum on a cylinder}

\begin{figure*}                                                       
\centerline{\includegraphics[width=2\columnwidth]{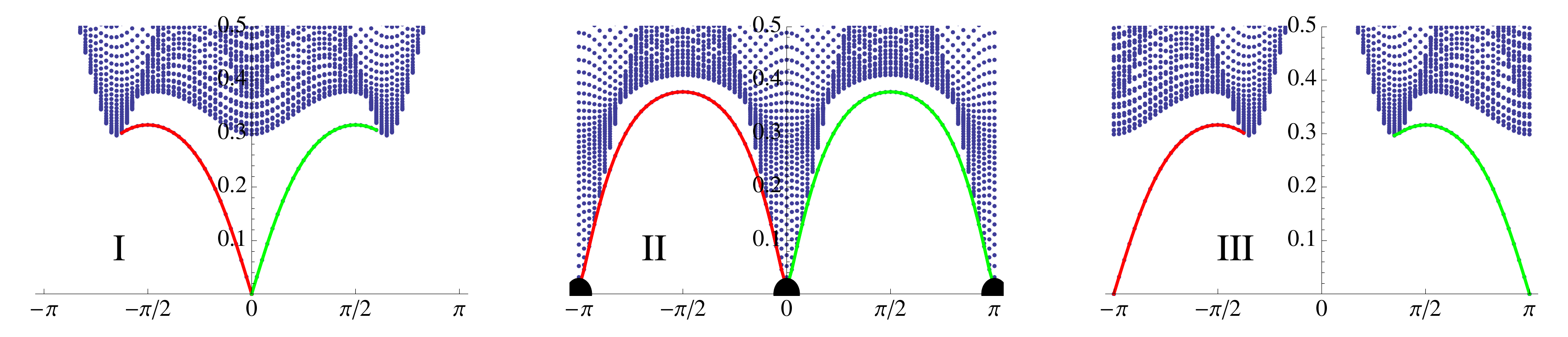}}
\caption{Spectrum on a cylinder, plotted as a function of $k_y$. In each figure, the gauge field strength $\beta$ and the imbalance $h$ are given by the labeled dots in fig.~\ref{topPhaseDiag}a.
The lines mark the edge states, and the thick dots mark the position of the gap-closing points. The CN of the upper band is 1 (-1) in the left (right) image, while the central image depicts a topological critical point.
The chemical potential is $\mu/(\sqrt{2}t)=-2$, and the energy on the vertical axis is in units of $t$.
}
\label{standardSpectrum}
\end{figure*}

\begin{figure*}
\centerline{\includegraphics[width=2\columnwidth]{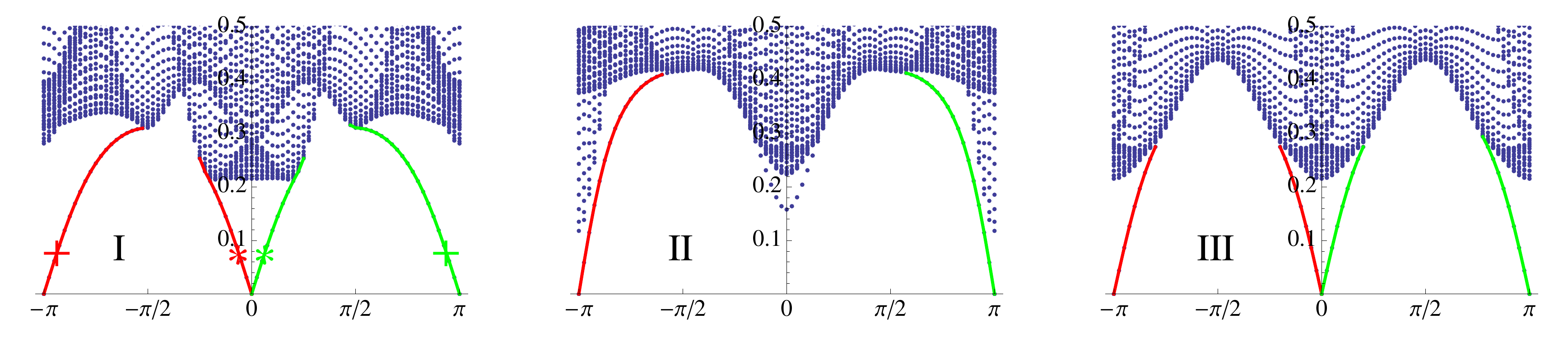}}
\caption{Same as fig.~\ref{standardSpectrum}, but here the chemical potential is $\mu/(\sqrt{2}t)=-0.5$, while the gauge  field strength $\beta$ and the imbalance $h$ correspond to the labeled dots in fig.~\ref{topPhaseDiag}c.
From left to right, the CN of the upper band is respectively -2,-1, and 0.
 The symbols + and * identify the eigenstates plotted in fig.~\ref{edges}.
}
\label{invertedSpectrum}
\end{figure*}

A characteristic feature of topological phases is the presence of intra-gap modes localized on open boundaries (edges) of the system. Ultracold gases with sharp edges may be realized along the lines proposed in ref.~\cite{goldmanEdges}. It should also be remembered that the core of a vortex is topologically equivalent to an edge which has been closed on itself. The topologically-protected edge modes we describe here are therefore equivalent to the Majorana fermions known to exist, e.g., in the core of vortices of p-wave superfluids.

To investigate this aspect, we calculate the spectrum of the Hamiltonian (\ref{completeHamiltonian}) on a cylinder, i.e., respectively with open and periodic boundary conditions along $x$ and $y$. Due to the broken translational symmetry along the $x$-direction, we diagonalize the Hamiltonian by performing a Fourier transformation along the $y$-direction only. 

\begin{figure}
\includegraphics[width=1.0\columnwidth]{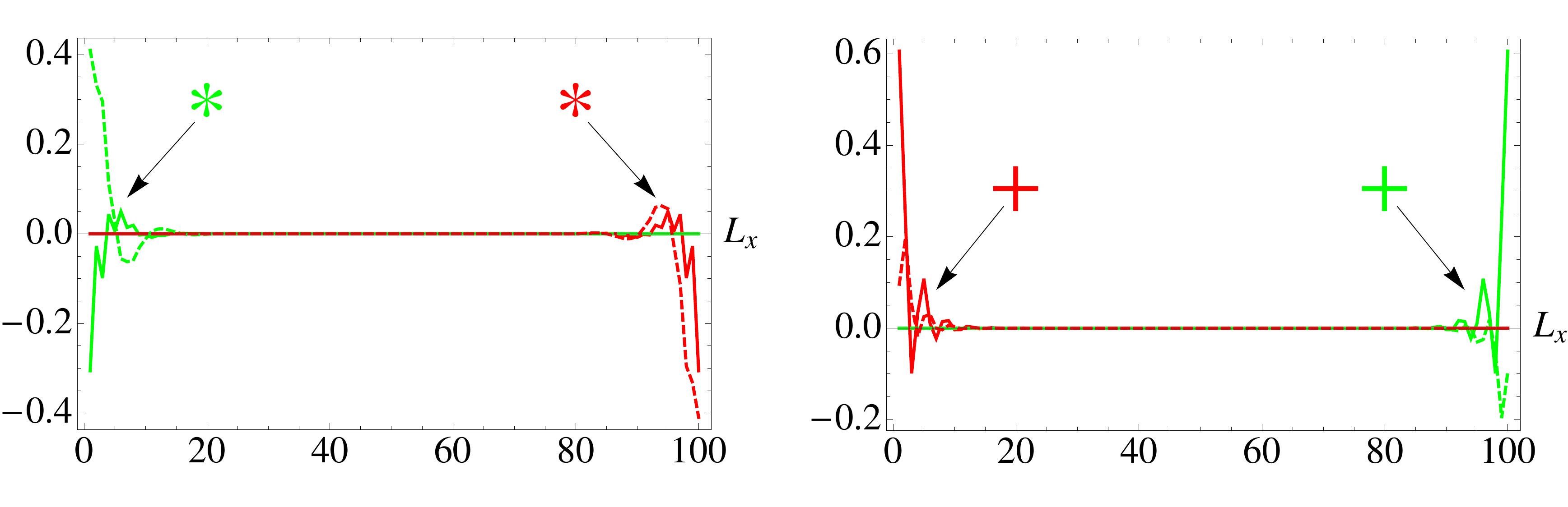}
\caption{Edge modes on a cylinder: continuous (dashed) lines depict the amplitude of the spin $\uparrow$ ($\downarrow$) component of the wavefunction with $\alpha,\beta, \mu$ and $h$ as in fig.~\ref{invertedSpectrum}(I).
 The momenta of the edge states are given by the symbols in fig.~\ref{invertedSpectrum}(I). Here two counterpropagating modes with identical spin state are localized on the same edge. In analogy with the QSH effect, these states are not topologically-protected, since a weak perturbation may induce backscattering from one mode to the other.
}
\label{edges}
\end{figure}

In figs.~\ref{standardSpectrum} and \ref{invertedSpectrum} we show the spectrum of $\mathcal{H}_{x,k_y}$ as a function of $k_y$ for various combinations of $\mu$, $\beta$, and $h$.
The spectrum on a torus would be generally gapped, but when open boundary conditions are taken into account isolated states appear inside the gap. The eigenvectors corresponding to these isolated eigenstates are linear superpositions of $\uparrow$ and $\downarrow$ spin components, exponentially localized on either of the boundaries, as may be seen in fig.~\ref{edges}. These are therefore termed "edge states". Edge modes in this context have first been discussed in ref.~\cite{sato} for  the specific case $\alpha=\beta=\pi/4$.

Given the reflection symmetry of the spectrum with respect to the sign of $k_y$, these states always appear in $\pm \tilde{k}$ pairs. Edge excitations may cross at zero energy with a linear dispersion at $k_y=0$ or $\pi$,  similarly to lattice Dirac fermions.
Their origin is topological, and these zero-energy crossings may appear or disappear only at the critical points $h=h_{\bk_0}$ where the bulk spectra of the upper and lower bands touch in zero, see e.g.\ fig.~\ref{standardSpectrum}(II).
Since the fully-balanced ($h=0$) and the fully-polarized ($h\gg t$) states are not topological and do not support any edge modes,  on a lattice we find that edge states are present only for imbalances satisfying $\mathrm{min}_{\bk_0} (h_{\bk_0})<h<\mathrm{max}_{\bk_0} (h_{\bk_0})$.

When a single pair of states is present, see e.g.\ figs.~\ref{standardSpectrum}a and \ref{standardSpectrum}c, the two counter-propagating edge modes with $\pm\tilde{k}$ are localized on opposite boundaries. Since on a given boundary there is no state available for backward spin-conserving scattering, these states are topologically protected.
 These conditions are realized when mod$_2$(CN)=1.

When instead the parity of CN is 0, as depicted in fig.~\ref{invertedSpectrum}, the system contains either zero or two pairs of edge states. At sufficiently low-energy, the eigenvectors at a given energy are characterized by symmetric quasi-momenta $k_y=\pm\tilde{k},\pm(\pi-\tilde{k})$. As may be seen in fig.~\ref{edges}, the counter-propagating edge states with $k_y=\tilde{k}$ and $k_y=-\pi+\tilde{k}$ are both localized on the same edge. When mod$_2$(CN)=0, we see that either edge modes can perform spin-conserving backscattering, or that there are no edge modes at all. A phase with even parity of the CN does not contain therefore any topologically-protected states.

It is interesting to note that fig.~\ref{invertedSpectrum}(III) depicts a physical state with two pairs of edge states but vanishing CN. To the best of our knowledge, this case has not been discussed so far in the context of cold atomic gases.

\section{Conclusions}

 We have studied here interacting ultracold fermions in presence of non-Abelian gauge fields. We have demonstrated that spin-imbalance can lead to a variety of intriguing topological phases. We characterized these phases by the associated Chern numbers, we presented the full phase diagram, and we discussed the link existing between edge states and Chern numbers. Furthermore, we have shown via a self-consistent calculation which of the phases can be realized experimentally, and which are non-physical due to pairing breakdown.
 In light of the recent ground-breaking experimental realizations of synthetic Abelian \cite{spielmanPRL,spielman, gemelke} fields in ultracold gases, and given the prospective to realize non-Abelian fields soon, the fascinating physics discussed in this manuscript should be readily accessible in the very close future.

\acknowledgments
We wish to thank G. Szirmai for a critical reading of the manuscript. We also acknowledge interesting discussions with M. Sato, S. Fujimoto, D. Poletti, S. Das Sarma, and N. Goldman.
We acknowledge funding from ESF/MEC project FERMIX (FIS2007-29996-E), Spanish MEC projects FIS2008-01236, TOQATA FIS2008-00784, QOIT (Consolider Ingenio 2010), Catalan project 2009-SGR-985, EU Integrated Project SCALA, EU STREP project NAMEQUAM, and ERC Advanced Grant QUAGATUA.
 A.K. acknowledges the Polish Government Scientific Funds 2009-2010, and M.L. the Humboldt Foundation Senior Research Prize.

\end{document}